\documentclass[aps,pra,twocolumn]{revtex4}%
\usepackage{amsfonts}
\usepackage{amsmath}
\usepackage{amssymb}
\usepackage{graphicx}
\usepackage{float}

\newcommand{\qed}{\nobreak \ifvmode \relax \else
\ifdim\lastskip<1.5em \hskip-\lastskip
\hskip1.5em plus0em minus0.5em \fi \nobreak
\vrule height0.75em width0.5em depth0.25em\fi}

\begin{document}
\title{Finite size analysis of measurement device independent\\ quantum cryptography
with continuous variables}
\author{Panagiotis Papanastasiou}
\affiliation{Computer Science and York Centre for Quantum Technologies, University of York,
York YO10 5GH, United Kingdom}
\author{Carlo Ottaviani}
\affiliation{Computer Science and York Centre for Quantum Technologies, University of York,
York YO10 5GH, United Kingdom}
\author{Stefano Pirandola}
\affiliation{Computer Science and York Centre for Quantum Technologies, University of York,
York YO10 5GH, United Kingdom}

\begin{abstract}
We study the impact of finite-size effects on the key rate of
continuous-variable (CV) measurement-device-independent (MDI)
quantum key distribution (QKD), considering two-mode Gaussian
attacks. Inspired by the parameter estimation technique developed
in [Ruppert \textit{et al.} Phys. Rev. A \textbf{90}, 062310
(2014)]~we adapt it to study CV-MDI-QKD and, assuming realistic
experimental conditions, we analyze the impact of finite-size
effects on the key rate. We find that the performance of the
protocol approaches the ideal one increasing the block-size and,
most importantly, that blocks between $10^{6}$ and $10^{9}$ data
points may provide key rates $\sim10^{-2}$ bit/use over
metropolitan distances.

\end{abstract}
\maketitle

\section{Introduction}

Quantum key distribution (QKD)~\cite{Gisin2002} promises to allow
unconditionally secure (theoretical) communication. Its strength relies on two
main elements: The encoding of classical information ($0$ and $1$ bits) into
non-orthogonal quantum states, and the impossibility of perfect discrimination
between them. In a conventional QKD protocol two users, Alice and Bob, share
quantum systems which are unavoidably corrupted every time that an
eavesdropper (Eve) tries to access the information encoded. This perturbation
is detectable, and allows the parties to quantify the amount of
error-correction and privacy amplification to apply to the shared data, in
order to reduce Eve's information to a negligible amount. Then they can use
the obtained key in an one-time pad protocol~\cite{schneider}.

The fundamental mechanism of QKD is clearly preserved also in more
complex (repeater-based) communication
configurations~\cite{Dur,DLCZ}, aiming at activating long distance
communication and quantum networks. In the basic point-to-point
scenario, the recent work~\cite{PLOB15} succeeded in establishing
the secret-key capacity of various quantum channels. The combined
use of relative entropy of entanglement~\cite{REE1,REE2,REE3} and
teleportation stretching (which reduces any adaptive protocol to a
block-form) enables one to compute the two-way capacity of many
important quantum channels (see Ref.~\cite{PLOB15} and further
works~\cite{Pirnetwork,multipoint,nonPauli,FiniteSTRET,PirCosmo}
for the correct definition and rigorous use of teleportation
stretching in quantum communication, quantum metrology and channel
discrimination). The result of Ref.~\cite{PLOB15} sets the
fundamental limit of point-to-point QKD and, as such, it marks the
edge when private communication inevitably needs quantum
repeaters. This benchmark has already had a wide application in
recent works~\cite{bench1,bench2,bench3,bench4,bench5,
3states,LoPiparo1,LoPiparo2,thermalLB16,JeffOSA, Bradler,
MihirRouting, Bash, LoPiparo1bis}.

Continuous variable (CV) quantum systems~\cite{RevModPhys.77.513}, in
particular Gaussian systems~\cite{Weedbrook2012}, emerged recently as very
promising carriers of quantum information. They have the potential to be used
for high-rate quantum communication because, rather than using single-photon
quantum states and photon-counting, they employ bright coherent states and
homodyne detections, which naturally boost the achievable key-rate. Based on
this premise, CV QKD protocols~\cite{diamantiENTROPY} have been proposed using
one-way~\cite{GG02,weedbrook2004noswitching,1D-usenko} or two-way quantum
communications~\cite{pirs2way}. Some one-way schemes have been experimentally
realized~\cite{Grosshans2003b,diamanti2007,1D-tobias, ulrik-Nat-Comm-2012},
over remarkably long-distance~\cite{jouguet2013,Huang-scirep2016}. Additional
theoretical analysis has been focused on QKD with thermal
states~\cite{filip-th1,weed1,usenkoTH1,weed2,weed2way,usenkoREVIEW}, with an
experiment performed~\cite{ulrik-entropy}. Recently CV-QKD has been extended
to a network configuration~\cite{CV-MDI-QKD,Ottaviani2015}, implementing the
general idea of measurement-device-independent (MDI) QKD~\cite{SamMDI,Lo}.
Here two parties, unable to access a secure direct link, can be assisted by an
intermediate relay (even untrusted) to establish a secure channel.

Many challenges~\cite{Scarani2008} need to be solved, before private quantum
networking can become a mature technology. However, CV QKD protocols and their
security analysis have progressed rapidly toward more practical and realistic
assumptions. In this respect, the incorporation of finite-size effects is
particularly important. In fact, when we assume that the parties exchange only
a finite number of signals, one expects the deterioration of the key rate. In
addition to this, finite-size analysis is also the first step towards a more
general security proof within the composable
framework~\cite{Furrer-PRL,leverrierCOMP,leverrierGEN}. While the theoretical
study of the impact of finite-size effects has been done in several previous
works~\cite{SCARANI-RENNER-PRL2008,Leverrierfsz,rupertPRA}, CV-MDI QKD has
been so far investigated neglecting this aspect and limiting the analysis to
the asymptotic regime~\cite{Devetak207}.

In order to start filling this gap, in this work we focus on
evaluating the impact of finite-size effects on the key-rate of
CV-MDI protocol. This study is important not only because these
effects have not yet evaluated in detail for MDI protocols, but
also because this type of analysis represents a necessary steps to
refine the security analysis of CV-MDI towards the more complete
composable scenario. We perform a detailed study of the impact of
finite-size effects for both the symmetric~\cite{Ottaviani2015}
and asymmetric configuration~\cite{CV-MDI-QKD,spedalieriSPIE}. We
extend the parameter estimation methods described in
Ref.~\cite{rupertPRA}, for conventional one-way protocols, to the
relay-based communication. We consider Gaussian two-mode attacks
which have been already extensively studied in one-way
schemes~\cite{1way2modes}, two-way
protocols~\cite{2way2modes,2way2modes2} and the CV-MDI
setup~\cite{two-modeNOTE}.

We remark that, in our analysis, we work within the Gaussian
assumption. This allows us to develop the statistical estimation
theory of the relevant parameters of the channels which are their
transmissivities and excess noise. The confidence interval of the
estimated parameters are then quantified using their variances and
setting a $6.5-$sigma accuracy, which allows us to grant a very
low error probability of $\epsilon_{PE}=10^{-10}$ during the
parameter estimation procedure. The confidence intervals are used
to select the worst-case scenario, by choosing the lower
transmissivity of the links and the higher excess-noise available.
The key rate is then numerically computed, using the estimated
values of transmissivity and noise, and optimized over free
parameters, which are the Gaussian modulation of the signals and
the ratio between the number of signals used in the parameter
estimation and total number of signals exchanged.

As expected we find that, increasing the block-size of the signals exchanged,
one recovers the performance under ideal condition. Most importantly, one has
that block-size in the range of $10^{6}\div10^{9}$ signals can provide a
positive key rate of about $10^{-2}$ bit/use, in the presence of high excess
noise $\varepsilon=0.01$, and attenuation compatible with the use of standard
optical fibres over metropolitan distances. The structure of this paper is the
following. In Section~\ref{The protocol}, we present the details of the
CV-MDI-QKD protocol. In Section~\ref{Channel parameter estimation}, we
describe the parameter estimation. In Section~\ref{Results}, we discuss the
results obtained and finally Section~\ref{conclusions} is for the conclusions.

\section{\label{The protocol}Protocol, eavesdropping and key rate}

For the sake of clarity, let us first describe the working mechanism of
CV-MDI-QKD from the prepare and measure perspective, where Alice and Bob send
coherent states, $|\alpha\rangle$ and $|\beta\rangle$, to an intermediate
relay. The amplitudes $\alpha$ and $\beta$ are Gaussian modulated, i.e., each
party sends to the relay an average thermal state with variance $V_{M}\geq0$.
The duty of the relay is very simple: it mixes the incoming signals on a
balanced beam splitter and performs a CV Bell detection, i.e., two conjugate
homodyne detections on $q_{-}$ and $p_{+}$, at the output ports of the
balanced beam splitter~\cite{bell}. \begin{figure}[ptb]
\centering\includegraphics[width=0.475\textwidth]{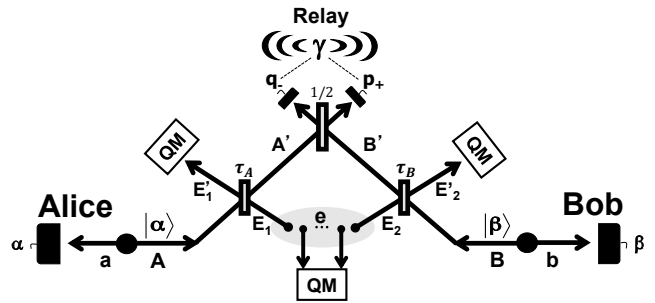}\caption{The
figure shows the EB representation of CV-MDI QKD. Alice and Bob have TMSV
states with modes $(a,A)$ and $(b,B)$. Local modes $a$ and $b$ are kept by the
parties, while $A$ and $B$ are sent to the relay through two links with
transmittance $\tau_{A}$ and $\tau_{B}$. When Alice and Bob heterodyne the
local modes, the travelling ones $A$ and $B$ are projected onto coherent
states $|\alpha\rangle$ and $|\beta\rangle$. The relay performs a
Bell-measurement and broadcast the outcomes $\gamma$, creating correlation
between the parties: for instance, Bob recovers Alice variable $\beta$ by
subtracting his variable $\alpha$ from the relay's outputs $\gamma$. The
Gaussian attack on the links is simulated by Eve using ancillas $E_{1}$ and
$E_{2}$, and thermal noise $\omega_{A}\geq1$ and $\omega_{B}\geq1$,
respectively. These ancillary modes are, in general, two-mode correlated (see
text for more details). The ancillary outputs are stored in a quantum memory
for a later measurement.}%
\label{relayprotocol}%
\end{figure}Then, the relay broadcasts the obtained values of $\gamma:=\left(
q_{-}+ip_{+}\right)  /\sqrt{2}$. This new variable can also be written as
$\gamma:=\alpha-\beta^{\ast}+\hat{\delta}$, where $\hat{\delta}$ is the
detection noise. It is then clear that the relay acts as a correlator for the
parties, who can infer each other variable ($\alpha$, $\beta$) from a simple
post-processing \cite{CV-MDI-QKD}.

The broadcast of the Bell detection outcomes, $\gamma$, does not help the
eavesdropper who is forced to attack the communication links to the relay, in
order to obtain information on amplitudes $\alpha$ and $\beta$. This operation
introduces detectable excess of noise, that the parties can use to quantify
Eve's knowledge on $\alpha$ and $\beta$ (accessible information). From this
stage on, the protocol works as any other QKD scheme \cite{Gisin2002}, with
the Alice and Bob implementing enough error correction and privacy
amplification to reduce Eve's accessible information to a negligible amount.

\subsection{Two-mode eavesdropping}

A powerful approach to study the security of any quantum cryptographic
protocol is to adopt the entanglement based (EB) representation, where the
description of the dynamics takes place in a dilated Hilbert space, which
allows to work with pure states. The EB representation of CV-MDI QKD scheme is
given in Fig.~\ref{relayprotocol}: Alice's and Bob's sources of coherent
states are purified assuming to start from a two-mode squeezed vacuum (TMSV)
states $\rho_{aA}$ and $\rho_{bB}$, whose zero-mean Gaussian states are
completely described by the following identical covariance matrices (CM)%
\begin{equation}
\mathbf{V}_{aA}=\mathbf{V}_{bB}=%
\begin{pmatrix}
\mu\mathbf{I} & \sqrt{\mu^{2}-1}\mathbf{Z}\\
\sqrt{\mu^{2}-1}\mathbf{Z} & \mu\mathbf{I}%
\end{pmatrix}
, \label{TMSV}%
\end{equation}
where $\mu=V_{M}+1$ and $\mathbf{Z=}$diag$(1,-1)$.

Modes $A$ and $B$ are sent through the links, while local ones, $a$ and $b$,
are heterodyned. The measurements projects the travelling modes into coherent
states $|\alpha\rangle$ and $|\beta\rangle$ respectively. The channel
attenuation on modes $A$ and $B$ is modeled by two beam splitters with
transmissivities $\tau_{A}$ and $\tau_{B}$, with $0\leq\tau_{A,B}\leq1$. These
process Alice's and Bob's signals with a pair of Eve's ancillary systems
$E_{1}$ and $E_{2}$ which, in general, belong to a wider reservoir of modes
controlled by the eavesdropper and including the set $\mathbf{e}$ (which can
be neglected in the limit of infinite signals exchanged~\cite{pirs1modeCh}).

We can then write the dilation of the initial Eve's state as a two-mode
Gaussian state $\sigma_{E_{1}E_{2}}$ described by the following general CM
\begin{equation}
\label{eq:attack}\mathbf{V}_{E_{1}E_{2}}=%
\begin{pmatrix}
\omega_{A}\mathbf{I} & \mathbf{G}\\
\mathbf{G} & \omega_{B}\mathbf{I}%
\end{pmatrix}
,
\end{equation}
where $\mathbf{G=}$diag($g,g^{\prime}$). The correlation parameters $g$ and
$g^{\prime}$ satisfy the constraints given in
Ref.~\cite{Entanglementreactivation}, while $\omega_{A},\omega_{B}\geq1$
account for the thermal noise injected by Eve, on each link, during the
attack. When $g=g^{\prime}=0$, the two-mode state $\sigma_{E_{1}E_{2}}$ is a
tensor product, which leads to a standard single-mode collective attack
realized by two independent entangling cloners~\cite{Grosshans2003b}. By
contrast for $g\neq0$ and $g^{\prime}\neq0$, the two entangling cloners are
not independent, and the optimal attack is two-mode coherent, as described in
\cite{CV-MDI-QKD,Ottaviani2015,two-modeNOTE}.

\subsection{Key rate}

The EB representation is useful for the security analysis because Alice-Bob
reduced output states $\rho_{ab|\gamma}$ and $\rho_{b|\gamma\alpha}$,
described by the CMs $\mathbf{V}_{ab|\gamma}$ and $\mathbf{V}_{b|\gamma\alpha
}$ respectively (see Section \ref{section-app-CM} for further details), have
the same entropies of Eve's output states. Under the ideal assumption that the
parties exchange infinite many signals ($N\gg1$), and assuming that the
parties reconcile over Alice's data to build the key, one bounds Eve's
accessible information by the Holevo function,
\begin{equation}
I_{H}:=S(\rho_{ab|\gamma})-S(\rho_{b|\gamma\alpha}), \label{holevo function}%
\end{equation}
where $S(.)$ is the von Neumann entropy. For Gaussian states, we have the
simple expression~\cite{Weedbrook2012}%
\[
S=\sum_{x}h(x),
\]
where $x$ is the generic symplectic eigenvalue of the CM, and
\begin{align}
h(x)  &  =\frac{x+1}{2}\log_{2}\frac{x+1}{2}-\frac{x-1}{2}\log_{2}\frac
{x-1}{2},\\
&  \overset{x\rightarrow\infty}{\rightarrow}\log_{2}\frac{e}{2}x.
\end{align}

We then can write an expression for the key rate
\begin{equation}
K^{\infty}:=\xi I_{AB}-I_{H}, \label{secret key rate}%
\end{equation}
where $\xi\leq1$ quantifies the inefficiency of error correction and privacy
amplification protocols \cite{jouguet2011,MILICEVIC,joseph} and $I_{AB}$ is
Alice-Bob mutual information. This is given by
\begin{equation}
I_{AB}=\frac{1}{2}\log_{2}\frac{V_{b|\gamma}^{q}+1}{V_{b|\gamma\alpha}^{q}%
+1}+\frac{1}{2}\log_{2}\frac{V_{b|\gamma}^{p}+1}{V_{b|\gamma\alpha}^{p}+1},
\label{mutual information}%
\end{equation}
with $V_{b|\gamma}^{q}$ ($V_{b|\gamma}^{p}$) and $V_{b|\gamma\alpha}^{q}$
($V_{b|\gamma\alpha}^{p}$) being the variances of CMs $\mathbf{V}_{ab|\gamma}$
and $\mathbf{V}_{b|\gamma\alpha}$ for the position (momentum) quadrature.
These CMs are given in Appendix \ref{section-app-CM}. The key rate is then
function of parameters $\xi$, $\omega_{A}$, $\omega_{B}$, $\tau_{A}$ and
$\tau_{B}$ and the Gaussian modulation $V_{M}$. Its expression can be found in
the supplemental information of Ref.~\cite{CV-MDI-QKD}.

\section{\label{Channel parameter estimation}Channel parameter estimation}

In a practical implementation of any QKD protocol, Alice and Bob can only
exchange a finite number of signals. In addition, they can only use a portion
of these to build the key, being the others used to estimate the channel
parameters. In this section we provide a description of CV-MDI QKD,
quantifying the impact of finite-size effects and the performance of the
protocol. To perform this analysis, we adapt the theory developed in
Ref.~\cite{rupertPRA} for one-way CV-QKD. We determine the channel parameters
(transmissivity and excess noise) within confidence intervals. Then we choose
the worst case scenario, picking the lower transmissivity and higher excess
noise within their confidence intervals, so as to minimize the key rate.

\subsection{Losses and excess noise at the relay outputs\label{lossandexcess}}

The outputs variables of the relay are quadratures $q_{-}$, relative to mode
$-$, and $p_{+}$ for mode $+$. These depend on the evolution of Alice's and
Bob's travelling modes $A=(q_{A}$, $p_{A})$ and $B=(q_{B}$, $p_{B})$. In terms
of these input field quadratures one can then write the following relations
\begin{align}
q_{-}  &  =\frac{1}{\sqrt{2}}(\sqrt{\tau_{B}}q_{B}-\sqrt{\tau_{A}}q_{A}%
)+q_{N},\label{q-}\\
p_{+}  &  =\frac{1}{\sqrt{2}}(\sqrt{\tau_{B}}p_{B}+\sqrt{\tau_{A}}p_{A}%
)+p_{N}, \label{p+}%
\end{align}
where $q_{N}=q_{\varepsilon}+q_{sn}$ and $p_{N}=q_{\varepsilon}+q_{sn}$ are
noise terms accounting for both excess noise and quantum shot noise coming
form the signal modes as well as Eve's ancillary modes. Their variances are
given by%
\begin{equation}
V_{q_{N}}=1+V_{q,\epsilon},\text{ \ \ }V_{p_{N}}=1+V_{p,\epsilon}, \label{VN}%
\end{equation}
with%
\begin{equation}
V_{q,\epsilon}=k-gu,\text{ \ \ }V_{p,\epsilon}=k+g^{\prime}u, \label{Veps}%
\end{equation}
and%
\begin{align}
k  &  =\frac{(1-\tau_{B})(\omega_{B}-1)+(1-\tau_{A})(\omega_{A}-1)}%
{2},\label{k}\\
u  &  =\sqrt{(1-\tau_{B})(1-\tau_{A})}, \label{u}%
\end{align}
where $g$ and $g^{\prime}$ have been defined in Eq.~(\ref{eq:attack}).

Now we describe in more detail the parameters estimation procedure. Alice's
and Bob's Gaussian modulation $V_{M}$ is assumed to be a known parameter. We
need to estimate the channel's transmissivity $\tau_{A}$, $\tau_{B}$ and
variance of the excess noises $V_{q,\epsilon}$ and $V_{p,\epsilon}$, with
their confidence intervals. Assuming that $m$ Gaussian distributed signals are
used for this task, we associate to $A_{q,i}$ ($A_{p,i}$) and $B_{q,i}$
($B_{p,i}$), for $i\in\{1,2,\dots,m\}$, the empirical realizations of the
field quadrature of the traveling modes. By contrast, we denote by $R_{q,i}$
and $R_{p,i}$ the realizations of the relay outputs. Let first discuss the
dynamics of the quadrature $q$. From Eq.~(\ref{q-}) one can write the
estimator of transmissivity $\tau_{A}$ as follows
\[
\hat{\tau}_{Aq}=\frac{2}{V_{M}^{2}}\hat{C}_{AR_{q}}^{2},
\]
where the covariance $C_{AR_{q}}=\sqrt{\tau_{A}/2}V_{M}$ has maximum
likelihood estimator given by%
\[
\hat{C}_{AR_{q}}=\frac{1}{m}\sum_{i=1}^{m}A_{q,i}R_{q,i},
\]
and to which one can associate the following variance (see
Appendix~\ref{The variances of the channel parameter estimators} for more
details)%
\begin{equation}
\text{Var}(\hat{\tau}_{Aq})=\frac{8\tau_{A}}{m}\left(  \tau_{A}+\frac{\tau
_{B}}{2}\right)  \left[  1+\frac{V_{q,N}}{\left(  \tau_{A}+\frac{\tau_{B}}%
{2}\right)  V_{M}}\right]  . \label{var-TA-q}%
\end{equation}
Very similar relations hold for the estimator of $\tau_{A}$ obtained
considering the other output of the relay, $p_{+}$. We can write the estimator
of the covariance $C_{AR_{p}}$, which is given by%
\[
\hat{C}_{ARp}=\frac{1}{m}\sum_{i=1}^{m}A_{p,i}R_{p,i},
\]
then using Eq.~(\ref{p+}) one can write the estimator of the transmissivity
$\tau_{A}$%
\[
\hat{\tau}_{Ap}=\frac{2}{V_{M}^{2}}\hat{C}_{AR_{p}}^{2},
\]
having variance
\begin{equation}
\text{Var}(\hat{\tau}_{Ap})=\frac{8}{m}\tau_{A}\left(  \tau_{A}+\frac{\tau
_{B}}{2}\right)  \left[  1+\frac{V_{p,N}}{\left(  \tau_{A}+\frac{\tau_{B}}%
{2}\right)  V_{M}}\right]  . \label{var-TA-p}%
\end{equation}
We notice that it differs from the formula of Eq.~(\ref{var-TA-q}) for the
expression of $V_{p,N}$, given in Eq.~(\ref{VN}). Now, from
Eq.~(\ref{var-TA-q}) and Eq.~(\ref{var-TA-p}) we calculate the optimum linear
combination of the variances of the two estimators
\begin{equation}
\text{Var}(\hat{\tau}_{A})=\frac{\text{Var}(\hat{\tau}_{A_{q}})\text{Var}%
(\hat{\tau}_{Ap})}{\text{Var}(\hat{\tau}_{A_{q}})+\text{Var}(\hat{\tau}_{Ap}%
)}:=\sigma_{A}^{2}. \label{sigA}%
\end{equation}
The same steps can be performed to obtain the relevant estimators for
transmissivity $\tau_{B}$ and the corresponding variance Var$(\hat{\tau}%
_{B})=\sigma_{B}^{2}$.

Now we write the estimator of the variance of the excess noise present on the
communication links, $V_{q,\epsilon}$. This can be derived from the maximum
likelihood estimator for $V_{q,N}$, and it reads
\begin{equation}
\hat{V}_{q,\epsilon}=\frac{1}{m}\sum_{i=1}^{m}\left[  R_{q,i}-\frac{1}%
{\sqrt{2}}(\sqrt{\hat{\tau}_{B}}B_{q,i}-\sqrt{\hat{\tau}_{A}}A_{q,i})\right]
^{2}-1, \label{VqEPS}%
\end{equation}
with variance
(Appendix~\ref{The variances of the channel parameter estimators})
\begin{equation}
\text{Var}(\hat{V}_{q,\epsilon})\approx\frac{2}{m}V_{q,N}^{2}:=s_{q}^{2}.
\label{varVqEPS}%
\end{equation}
Correspondingly, we obtain an estimator for $V_{p,\epsilon}$ expressed as
\begin{equation}
\hat{V}_{p,\epsilon}=\frac{1}{m}\sum_{i=1}\left[  R_{p,i}-\frac{1}{\sqrt{2}%
}(\sqrt{\hat{\tau}_{B}}B_{p,i}+\sqrt{\hat{\tau}_{A}}A_{p,i})\right]  ^{2}-1
\label{VpEPS}%
\end{equation}
and variance
\begin{equation}
\text{Var}(\hat{V}_{p,\epsilon})\approx\frac{2}{m}V_{p,N}^{2}:=s_{p}^{2}.
\label{varVpEPS}%
\end{equation}
Finally, from Eq.~(\ref{VqEPS}) and Eq.~(\ref{varVpEPS}) we compute the
confidence intervals and select the pessimistic values given by the following
choice of parameters%
\begin{align}
\tau_{A}^{\text{\textrm{low}}}=  &  \tau_{A}-6.5\sigma_{A},\text{ \ \ }%
\tau_{B}^{\text{\textrm{low}}}=\tau_{B}-6.5\sigma_{B},\label{taulow}\\
V_{q,\epsilon}^{\text{\textrm{up}}}=  &  V_{q,\epsilon}+6.5s_{q},\text{
\ \ }V_{p,\epsilon}^{\text{\textrm{up}}}=V_{p,\epsilon}+6.5s_{p}. \label{Vup}%
\end{align}

\subsection{Secret key rate with finite size effects}

Once we have obtained the estimation of the transmissivities of the links and
the corresponding excess noises, we can write the key rate, incorporating
finite size effects writing%
\begin{equation}
K=\frac{n}{\bar{N}}\left(  K^{\infty}(\xi,V_{M},\tau_{A}^{\text{\textrm{low}}%
},\tau_{B}^{\text{\textrm{low}}},V_{q,\epsilon}^{\text{\textrm{up}}%
},V_{p,\epsilon}^{\text{\textrm{up}}})-\Delta(n)\right)  ,
\label{K-finitesizeeffects}%
\end{equation}
where $n=\bar{N}-m$ is the number of signals used to prepare the key and
$\bar{N}$ the total number of signal exchanged. The key rate is then computed
replacing the values of Eq.~(\ref{taulow}) and Eq.~(\ref{Vup}) in the
asymptotic key rate of Eq.~(\ref{secret key rate}). In particular, first one
computes the key rate $R$ of Eq.~(\ref{K-finitesizeeffects}), using the Holevo
function of Eq.~(\ref{holevo function}) for the channel parameters given by
Eq.~(\ref{taulow}) and (\ref{Vup}). Then, in order \begin{figure}[ptb]
\includegraphics[width=0.55\textwidth]{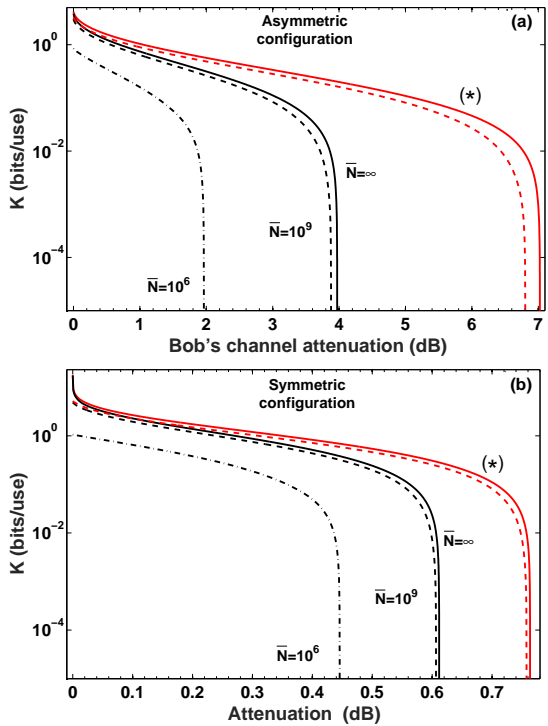}\vspace{-0.25cm}%
\caption{(Color online)The figure summarize the impact of finite size effects
on the performance of CV-MDI QKD for both asymmetric (panel a) and symmetric
(panel b) configuration of the relay, in the presence of optimal two-mode
attack. In panel (a) the key rate is plotted as a function of the dB of
attenuation on Bob's channel, with the relay placed near Alice $\tau_{A}%
=0.98$. From top to bottom, the black curves describe: the rate for $\bar
{N}\gg1$ with $\xi=0.98$, and optimizing over $V_{M}$ (solid line). Then we
have the cases with finite block-size. The dashed line is for $\bar{N}=10^{9}$
while the dot-dashed curve is obtained for $\bar{N}=10^{6}$. In all cases we
have assumed excess of noise $\varepsilon=0.01$ SNU. The red curves (*)
describe the case obtained for pure loss and assuming $\xi=1$, $V_{M}%
\rightarrow\infty$, $\bar{N}\rightarrow\infty$ (solid line) and $\bar
{N}=10^{9}$ \ $\ $(dashed line). Panel (b) focuses on the symmetric
configuration of the relay. The curves are obtained using the same parameters
as in panel (a), but setting $\tau_{A}=\tau_{B}=\tau$.}%
\label{FIG2}%
\end{figure}to account for the penalty for using the Holevo function even if
we have a finite number of signals exchanged, one must include the correction
term%
\[
\Delta(n)\sim\sqrt{\frac{1}{n}\log_{2}2\varepsilon_{PA}^{-1}}%
\]
which depends on the number of signals used to prepare the key, $n$, and the
probability of error related to the privacy amplification procedure
$\varepsilon_{PA}$. A detailed description of this correction term can be
found in Ref.~\cite{Leverrierfsz}.

\section{\label{Results} Results}

\begin{figure}[ptb]
\includegraphics[width=0.45\textwidth]{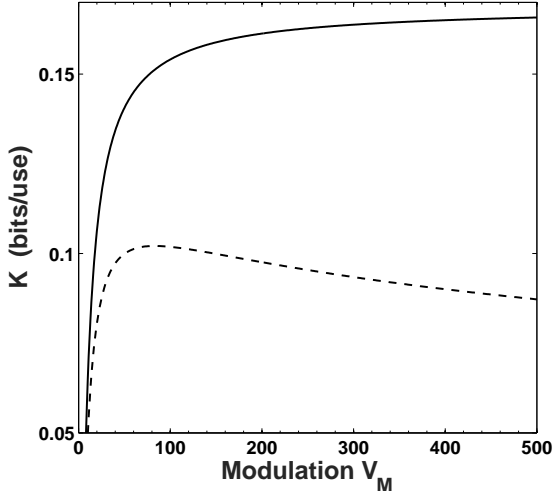}\vspace{-0.25cm}%
\caption{This figure shows the impact of the reconciliation efficiency on the
key-rate. When $\xi=0.95$ (dashed line), the Gaussian modulation maximizing
the key rate is $V_{M}<\infty$ . While when $\xi=1$ then the optimal key rate
is obtained for $V_{M}\rightarrow\infty$ \ (solid line). The lines are
obtained for pure loss attack, $\tau_{A}=0.98$ and $\tau_{B}=0.7$ and block
size $\bar{N}=10^{6}$.}%
\label{R-vs-VM}%
\end{figure}

The key rate of the asymmetric configuration of the relay is described in the
top-panel (a) of Fig.~\ref{FIG2}. We plot the key rate as a function of Bob's
channel transmissivity, expressed in terms of dB of attenuation, while \ the
transmissivity of Alice's link is set to $\tau_{A}=0.98$. The curves are
obtained considering two-mode optimal attacks, for which $g=-g^{\prime}$ with
$g=\min\left[  \sqrt{(\omega_{A}-1)(\omega_{B}+1)},\sqrt{(\omega_{B}%
-1)(\omega_{A}+1)})\right]  $ and $\omega_{A}\sim\omega_{B}\sim1.01$%
~\cite{CV-MDI-QKD} and using the key rate of Eq.~(\ref{K-finitesizeeffects})
incorporating also finite-size effects. The efficiency of classical code for
error correction and reconciliation efficiency is set to $\xi=0.98$, and the
final key rate is optimized over the variance of the Gaussian modulation
$V_{M}$ (see Fig.~\ref{R-vs-VM}) and the ratio $r=n/\bar{N}$. The black-solid
line gives the asymptotic key rate for very large $\bar{N}$ ($>>10^{9}$),
while the dashed line is for block-size $\bar{N}=10^{9}$ and the dot-dashed
line is obtained for $\bar{N}=10^{6}$.

The bottom panel in Fig.~\ref{FIG2} (b), plots the secret key rate for the
symmetric case \cite{Ottaviani2015} ($\tau_{A}=\tau_{B}$). The curves are
obtained setting all the other parameters as in Fig.~\ref{FIG2}~(a) and
optimizing as before the key rate for the case including finite-size effects.
The black solid line describes again the asymptotic case $\bar{N}%
\rightarrow\infty$ of the symmetric configuration while the dashed lines is
obtained for $\bar{N}=10^{9}$ and the dotted one for $\bar{N}=10^{6}$.

Let us finally remark on a couple of points. First, we notice that the
performance of finite-size CV-MDI-QKD converges to the ideal one if the number
of signal exchanged increases. Second, according to the plots, we notice that
the key rate of the CV-MDI-QKD protocol is sufficiently robust with respect to
the finite size effects, with block sizes of $10^{9}$ points approaching the
asymptotic limit.

\section{Conclusion\label{conclusions}}

In conclusion, we have studied the security of Gaussian CV-MDI-QKD taking into
account finite-size effects. These emerge when one assumes that the parties
exchange only a finite number of signals during the quantum communication
stage. In our analysis we assumed imperfect efficiency of error correction and
privacy amplification $(\xi<1)$ and developed the finite-size analysis
adapting the channel parameters estimation approach described in
Ref.~\cite{rupertPRA}. The resulting finite size key rate has then been
optimized over the Gaussian modulation and the number of signals used to
perform the parameter estimation.

Our results show that, also considering finite-size effects under realistic
conditions, CV-MDI QKD over metropolitan distances is feasible within today
state-of-the-art experimental conditions. In particular, we found that the
adoption of block-size in the range $\bar{N}=10^{6}\div10^{9}$ is already
sufficient in order to achieve a high key rate of $10^{-2}$ bits/use over
metropolitan distances, and in the presence of an excess of noise
$\varepsilon=0.01$ (SNU).

Finally we underline that the present analysis does not put a final word on
the performances of finite-size CV-MDI-QKD. The validity of the described
analysis is in fact restricted to the case of Gaussian attacks. Further
studies are needed where finite-size effects are investigated within the
composable security framework. See Ref.~\cite{cosmoMDI-COMP} for a fully composable security
proof of CV-MDI-QKD.

\textit{Note added.} Shortly after the submission of our
manuscript, an independent work~\cite{JO-FS} has been posted on
the arXiv. Also this work studies the impact of finite-size blocks
on the key-rate of CV-MDI QKD.

\section{Acknowledgements}
C.O. acknowledges Cosmo Lupo for useful discussions. This work has
been supported by the EPSRC via the `UK Quantum Communications
HUB' (Grant no. EP/M013472/1).


\appendix

\section{ covariance matrices and symplectic eigenvalues\label{section-app-CM}%
}

For the sake of clarity, in this section we re-write the CMs and the relevant
symplectic eigenvalues derived in Ref.~\cite{CV-MDI-QKD} using the notation
adopted in the main text. The CM describing the total output state of Alice
and Bob, after the relay measurements, $\rho_{ab|\gamma}$, is given by the
following expression%

\begin{align}
\mathbf{V}_{ab|\gamma}  &  =%
\begin{pmatrix}
(V_{M}+1)\mathbf{I} & 0\\
0 & (V_{M}+1)\mathbf{I}%
\end{pmatrix}
-V_{M}(V_{M}+2)\times\nonumber\\
&  \times%
\begin{pmatrix}
\frac{\tau_{A}}{\varphi} & 0 & -\frac{\sqrt{\tau_{A}\tau_{B}}}{\varphi} & 0\\
0 & \frac{\tau_{A}}{\varphi^{\prime}} & 0 & \frac{\sqrt{\tau_{A}\tau_{B}}%
}{\varphi^{\prime}}\\
-\frac{\sqrt{\tau_{A}\tau_{B}}}{\varphi} & 0 & \frac{\tau_{B}}{\varphi} & 0\\
0 & \frac{\sqrt{\tau_{A}\tau_{B}}}{\varphi^{\prime}} & 0 & \frac{\tau_{B}%
}{\varphi^{\prime}}%
\end{pmatrix}
,
\end{align}
where
\begin{align}
\varphi &  =(\tau_{A}+\tau_{B})V_{M}+2+2V_{\epsilon,g},\\
\varphi^{\prime}  &  =(\tau_{A}+\tau_{B})V_{M}+2+2V_{\epsilon,g^{\prime}}%
\end{align}
and where we have defined
\begin{align}
V_{\epsilon,g}=  &  \frac{1}{2}\left(  \bar{\tau}_{B}(\omega_{B}-1)+\bar{\tau
}_{A}(\omega_{A}-1)-2g\sqrt{\bar{\tau}_{B}\bar{\tau}_{A}}\right)
,\label{eq:veg}\\
V_{\epsilon,g^{\prime}}=  &  \frac{1}{2}\left(  \bar{\tau}_{B}(\omega
_{B}-1)+\bar{\tau}_{A}(\omega_{A}-1)+2g^{\prime}\sqrt{\bar{\tau}_{B}\bar{\tau
}_{A}}\right)  , \label{eq:vegp}%
\end{align}
with $\bar{\tau}_{l}=1-\tau_{l}$, for $l=A,B$.

Bob's output CM after the double conditioning, first with respect the relay
measurements and then after Alice's heterodyne detection, $\mathbf{V}%
_{ab|\gamma}$, is given by
\[
\mathbf{V}_{b|\gamma\alpha}=%
\begin{pmatrix}
\frac{2(V_{M}+1)(1+V_{\epsilon,g})-\tau_{B}V_{M}}{2(1+V_{\epsilon,g})+\tau
_{B}V_{M}} & 0\\
0 & \frac{2(V_{M}+1)(1+V_{\epsilon,g^{\prime}})-\tau_{B}V_{M}}{2(1+V_{\epsilon
,g^{\prime}})+\tau_{B}V_{M}}.
\end{pmatrix}
\]
which has the following symplectic eigenvalue given by the $\sqrt{.}$ of the
determinant of previous matrix%
\begin{equation}
\bar{\nu}=\sqrt{\det\mathbf{V}_{b|\gamma\alpha}}.
\end{equation}

\section{Useful elements of estimation theory}

According to the method of maximum likelihood estimation, for a bivariate
normal distribution $X=(X_{1},X_{2}$), the estimators for the mean $\mu
=(\mu_{1},\mu_{2})$ and the covariance matrix $\mathbf{V}$ are given by%
\begin{align}
\hat{\boldsymbol{\mu}}=  &  \frac{1}{m}\sum_{i=1}^{m}\mathbf{X}_{i}\\
\widehat{\mathbf{V}}=  &  \frac{1}{m}\sum_{i=1}^{m}(\mathbf{X}_{i}%
-\hat{\boldsymbol{\mu}})(\mathbf{X}_{i}-\hat{\boldsymbol{\mu}})^{T}%
\end{align}
where $\mathbf{X}_{i}$ is the $i$-th statistical realization out of $m$
realizations of $\mathbf{X}$.

The {central limit theorem states that,} assuming $m$ realizations
$X_{1},X_{2},\dots,X_{m}$ ($m\gg1$) of a random variable $X$ with unknown
density function $f$, mean $\mu$ and variance $\sigma^{2}<\infty$, the sample
mean
\begin{equation}
\bar{X}=\dfrac{1}{m}\sum_{i=1}^{m}X_{i}%
\end{equation}
is approximately normal with mean $\mu$ and variance $\sigma^{2}/m$. In order
to estimate the mean value of a variable $Y$, that depends on the square of a
variable $X$ for which we have $m$ realizations, we can use the following
result: For $m$ realizations $X_{i}$, for $i=1,2,\dots,m$, of a normally
distributed variable $X$, having mean $\mu$ and unit variance, the variable
\begin{equation}
Y=\sum_{i=1}^{m}X_{i}^{2}\sim\chi^{2}\left(  k,\lambda\right)
\end{equation}
is distributed according to the $\chi^{2}$ distribution with $k=m$ degrees of
freedom and $\lambda=m\mu^{2}$. The mean value and variance of the chi-squared
distribution is given by
\begin{equation}
\mathbb{E}(Y)=k+\lambda,
\end{equation}
and%
\begin{equation}
\text{Var}(Y)=2(k+2\lambda).
\end{equation}

Let us assume to have two estimators $\hat{s}_{1}$ and $\hat{s}_{2}$, with
variances $\sigma_{1}^{2}$ and $\sigma_{2}^{2}$, for the same quantity $s$
acquired by different processes. We then compute the optimal linear
combination of the variances by the following formula%
\begin{equation}
\sigma_{\text{\textrm{opt}}}^{2}=\frac{\sigma_{1}^{2}\sigma_{2}^{2}}%
{\sigma_{1}^{2}+\sigma_{2}^{2}}.
\end{equation}

\section{Variances of the channel parameter
estimators\label{The variances of the channel parameter estimators}}

Let us suppose that $A_{q,i}$ ($B_{q,i}$) are independent variables each one
following the normal distribution $q_{A}$ ($q_{B}$) with zero mean, variance
$V_{M}$ as described in \ref{lossandexcess}. Accordingly, $R_{q,i}$ ($R_{p,i}%
$) are assumed to be independent variables following the normal distribution
of $q_{R}$ ($p_{R}$), i.e.\ the relay output variable.

\subsection{Variance of the transmissivity}

For the covariance between modes $A$ and $R$, we can write the following
estimator%
\begin{equation}
\hat{C}_{AR_{q}}=\frac{1}{m}\sum_{i=1}^{m}A_{q,i}R_{q,i}%
\end{equation}
normally distributed as the sample mean of variable $Z=A_{q}R_{q}$. We can
compute the expectation value by%
\begin{equation}
\mathbb{E}(\hat{C}_{AR_{q}})=\mathbb{E}(q_{A}q_{R})=\sqrt{\frac{\tau_{A}}{2}%
}V_{M}=C_{AR_{q}}, \label{covarianceARq}%
\end{equation}
and the variance can be defined as
\begin{equation}
V_{\text{\textrm{Cov}}}:=\text{\textrm{Var}}(\hat{C}_{AR_{q}})
\end{equation}
with
\begin{align}
\text{\textrm{Var}}(\hat{C}_{AR_{q}})  &  =\frac{1}{m}\text{\textrm{Var}%
}(q_{A}q_{R})=\nonumber\\
&  =\frac{1}{2m}\left[  \tau_{A}\text{\textrm{Var}}(q_{A}^{2})+\tau
_{B}\text{\textrm{Var}}(q_{A}q_{A,B})\right] \nonumber\\
&  +\text{\textrm{Var}}(q_{A}q_{N}),\\
&  =\frac{1}{m}\left(  \tau_{A}V_{M}^{2}+\frac{\tau_{B}}{2}V_{M}^{2}%
+V_{M}V_{q,N}\right) \nonumber\\
&  =\frac{1}{m}\left(  \tau_{A}+\frac{\tau_{B}}{2}\right)  V_{M}^{2}\left[
1+\frac{V_{q,N}}{\left(  \tau_{A}+\frac{\tau_{B}}{2}\right)  V_{M}}\right]  ,
\label{varianceofcovarianceARq}%
\end{align}
where we have considered the independence of the variables and the second
order moments of the normal distribution.

Therefore, we can derive the mean and variance for the estimator of $\tau_{A}%
$. We rewrite the estimator as
\begin{equation}
\hat{\tau}_{Aq}=\frac{2V_{\text{\textrm{Cov}}}}{V_{M}^{2}}\left(  \frac
{\hat{C}_{AR_{q}}}{\sqrt{V_{\text{\textrm{Cov}}}}}\right)  ^{2}.
\end{equation}
Note that the variable $\left(  \hat{C}_{AR_{q}}/\sqrt{V_{\text{\textrm{Cov}}%
}}\right)  ^{2}$ is $\chi^{2}-$distributed, i.e.,%
\begin{equation}
\left(  \frac{\hat{C}_{AR_{q}}}{\sqrt{V_{\text{\textrm{Cov}}}}}\right)
^{2}\sim\chi^{2}\left[  1,\left(  \frac{C_{AR_{q}}}{\sqrt
{V_{\text{\textrm{Cov}}}}}\right)  ^{2}\right]  ,
\end{equation}
with expectation value%
\begin{equation}
\mathbb{E}\left[  \left(  \frac{\hat{C}_{AR_{q}}}{\sqrt{V_{\text{\textrm{Cov}%
}}}}\right)  ^{2}\right]  =1+\left(  \frac{C_{AR_{q}}}{\sqrt
{V_{\text{\textrm{Cov}}}}}\right)  ^{2},
\end{equation}
and variance%
\begin{equation}
\text{Var}\left[  \left(  \frac{\hat{C}_{AR_{q}}}{\sqrt{V_{\text{\textrm{Cov}%
}}}}\right)  ^{2}\right]  =2\left[  1+2\left(  \frac{C_{AR_{q}}}%
{\sqrt{V_{\text{\textrm{Cov}}}}}\right)  ^{2}\right]  .
\end{equation}
The expectation value of $\hat{\tau}_{Aq}$ is then given by
\begin{align}
\mathbb{E}(\hat{\tau}_{Aq})  &  =\frac{2V_{\text{\textrm{Cov}}}}{V_{M}^{2}%
}\left[  1+\left(  \frac{C_{AR_{q}}}{\sqrt{V_{\text{\textrm{Cov}}}}}\right)
^{2}\right]  =\nonumber\\
&  =\frac{2C_{AR_{q}}^{2}}{V_{M}^{2}}+\mathcal{O}(1/m)=\tau_{A}+\mathcal{O}%
(1/m)
\end{align}
and its variance is%
\begin{align}
\text{Var}(\hat{\tau}_{Aq})  &  =\frac{4V_{\text{Cov}}^{2}}{V_{M}^{4}}2\left[
1+2\left(  \frac{C_{AR_{q}}}{\sqrt{V_{\text{Cov}}}}\right)  ^{2}\right]
=\nonumber\\
&  =\frac{16V_{\text{Cov}}C_{AR_{q}}^{2}}{V_{M}^{4}}+\mathcal{O}(1/m^{2}).
\end{align}

By replacing from Eq.~(\ref{covarianceARq}) and
Eq.~(\ref{varianceofcovarianceARq}), we obtain
\begin{align}
\text{Var}(\hat{\tau}_{Aq})  &  =\frac{16}{mV_{M}^{4}}\left(  \tau_{A}%
+\frac{\tau_{B}}{2}\right) \nonumber\\
&  \frac{V_{M}^{4}\tau_{A}}{2}\left[  1+\frac{V_{q,N}}{\left(  \tau_{A}%
+\frac{\tau_{B}}{2}\right)  V_{M}}\right]  +\mathcal{O}(1/m^{2}),\\
&  =\frac{8\tau_{A}}{m}\left(  \tau_{A}+\frac{\tau_{B}}{2}\right)  \left[
1+\frac{V_{q,N}}{\left(  \tau_{A}+\frac{\tau_{B}}{2}\right)  V_{M}}\right]
\nonumber\\
&  +\mathcal{O}(1/m^{2}).
\end{align}
Clearly, as previously noted in~\cite{rupertPRA}, the estimator of the
transmissivity $\hat{\tau}_{Aq}$ is only asymptotically unbiased. In fact the
standard deviation $\sqrt{\text{Var}(\hat{\tau}_{Aq})}$ is of order
$1/\sqrt{m}$ while the bias goes as $1/m$. As we consider $m>10^{5}$ in our
analysis, the value of the bias become rapidly negligible as $m\gg1$, and the
use of estimators $\hat{\tau}_{A}$ and $\hat{\tau}_{B}$ very accurate.

\subsection{Variance of the excess noise}

Also in our estimation procedure for the MDI protocol, in analogy to the
theory developed in Ref.~\cite{rupertPRA}, the variance of the estimator can
be obtained replacing the estimator of $\tau_{A}$ ($\tau_{B}$) with its value.
This simplifies the calculations. Now, we can assume that any uncertainty in
the estimator of the excess noise obtained from broadcast results of relay's
measurements on quadrature $q$%
\begin{equation}
\hat{V}_{q,\epsilon}=\frac{1}{m}\sum_{i=1}^{m}\left[  R_{q,i}-\frac{\sqrt
{\hat{\tau}_{B}}B_{q,i}-\sqrt{\hat{\tau}_{A}}A_{q,i}}{\sqrt{2}}\right]
^{2}-1, \label{excessnoisevar}%
\end{equation}
comes only from variables $R_{q,i}$, $A_{q,i}$ and $B_{q,i}$. We then have
that the expression inside square brackets is normally distributed with zero
mean and variance $V_{q,N}$. In addition to this one also has that the
following expression%
\begin{equation}
Y:=\sum_{i=1}^{m}\left(  \frac{R_{q,i}-\left(  \sqrt{\tau_{B}}B_{q,i}%
-\sqrt{\tau_{A}}A_{q,i}\right)  /\sqrt{2}}{\sqrt{V_{q,N}}}\right)  ^{2}%
\sim\chi^{2}(m,0),
\end{equation}
is $\chi^{2}$-distributed, and has mean $\mathbb{E}(Y)=m$ and variance
$\text{Var}(Y)=2m$. This allows to approximate the sum of
Eq.~(\ref{excessnoisevar}) with $V_{q,N}Y$ when we assume large values for
$m$, obtaining the expectation value%
\begin{equation}
\mathbb{E}(\hat{V}_{q,\epsilon})\approx\frac{1}{m}V_{q,N}\mathbb{E}%
(Y)-1=V_{q,\epsilon}%
\end{equation}
and the variance
\begin{equation}
\text{Var}(\hat{V}_{q,\epsilon})\approx\frac{1}{m^{2}}V_{q,N}^{2}%
\text{Var}(Y)=\frac{2}{m}V_{q,N}^{2}.
\end{equation}

\end{document}